\pdfoutput=1
\documentclass[bibyear,traditabstract]{aa}  
\usepackage{graphicx}
\usepackage{longtable}
\usepackage[varg]{txfonts}
\begin{document}
\title{Inversion of stellar fundamental parameters from Espadons and
  Narval high-resolution spectra\thanks{Based on observations obtained
    at the \emph{T\'elescope Bernard Lyot} (TBL, \emph{Pic du Midi},
    France), which is operated by the \emph{Observatoire
      Midi-Pyr\'en\'ees, Universit\'e de Toulouse, Centre National de
      la Recherche Scientifique} (France) and the Canada-France-Hawaii
    Telescope (CFHT) which is operated by the National Research Council of
    Canada, CNRS/INSU and the University of Hawaii (USA).}  }

\author{F. Paletou \and T. B\"ohm \and V. Watson \and J.-F. Trouilhet}

          \institute{Universit\'e de Toulouse, UPS-Observatoire
            Midi-Pyr\'en\'ees, IRAP, Toulouse, France\\
\and
CNRS,  Institut de Recherche en Astrophysique et
            Plan\'etologie, 14 av. E. Belin, F--31400 Toulouse, France\\
            \email{fpaletou@irap.omp.eu}
}

   \date{Received August 4, 2014; accepted November 10, 2014}

% \abstract{}{}{}{}{} 
% 5 {} token are mandatory

   \abstract {The general context of this study is the inversion of
     stellar fundamental parameters from high-resolution Echelle
     spectra.  We aim indeed at developing a fast and reliable tool
     for the post-processing of spectra produced by Espadons and
     Narval spectropolarimeters.  Our inversion tool relies on
     principal component analysis. It allows reduction of
     dimensionality and the definition of a specific metric for the
     search of nearest neighbours between an observed spectrum and a
     set of observed spectra taken from the Elodie stellar
     library. Effective temperature, surface gravity, total
     metallicity and projected rotational velocity are derived.
     Various tests presented in this study, and done from the sole
     information coming from a spectral band centered around the
     Mg\,{\sc i} b-triplet and with spectra from FGK stars are very
     promising.}


   \titlerunning{Inversion of stellar fundamental parameters from
     Espadons and Narval spectra.} 

   \maketitle

%
%________________________________________________________________

\section{Introduction}

This study is concerned with the inversion of fundamental stellar
parameters from the analysis of high-resolution Echelle spectra.
Hereafter, we shall focus indeed on data collected since 2006 with the
Narval spectropolarimeter mounted at the 2-m aperture
\emph{T\'elescope Bernard Lyot} (TBL) telescope located at the summit
of the \emph{Pic du Midi de Bigorre} (France) and, since 2005 with the
Espadons spectropolarimeter mounted at the 3.6-m aperture CFHT
telescope (Hawaii). We investigate, in particular, the capabilities of
the principal component analysis (hereafter PCA) for setting-up a fast
and reliable tool for the inversion of stellar fundamental parameters
from these high-resolution spectra.

The inversion of stellar fundamental parameters for each target that
was observed with both Narval and Espadons spectropolarimeters
constitutes an essential step towards: \emph{(i)} the further
post-processing of the data like e.g., the extraction of polarimetric
signals (see e.g., Paletou 2012 and references therein) but also
\emph{(ii)} the exploration, or data mining, of the full set of data
accumulated over the last 8 years now. In Section 2, we briefly
describe the actual content of such a database\footnote{ {\tt
    polarbase.irap.omp.eu} }.

PCA have been used for stellar spectral classification since Deeming
(1964). It has been in use since, and more recently for the purpose of
the \emph{inversion} of stellar fundamental parameters from the
analysis of spectra of various resolutions. It is however most often
used \emph{together} with artificial neural networks (see e.g.,
Bailer-Jones 2000; Re Fiorentin et al. 2007). PCA is used there for
reducing the dimensionality of the spectra before attacking a
multi-layer perceptron which, in turns, allows to link input data to
stellar parameters.

Our usage of PCA for such an inversion process is strongly influenced
by the one routinely made in \emph{solar} spectropolarimetry during
the last decade after the pionneering work of Rees et al. (2000). Very
briefly, the reduction of dimensionality allowed by PCA is directly
used for building a specific metric from which a nearest neighbour(s)
search is done between an observed data set and the content of a
``training database''. The latter can be made of synthetic data
generated from input parameters properly covering the \emph{a priori}
range of physical parameters expected to be deduced from the
observations themselves. A quite similar use of PCA was also presented
for classification and redshift of galaxies estimation by Cabanac et
al. (2002). However, in this study we shall use \emph{observed}
spectra from the Elodie stellar library, as our data set of reference
(Prugniel et al. 2007). Such spectra have been, for instance, used in
a recent study related to the determination of atmospheric parameters
of FGKM stars in the Kepler field (Molenda-{\.Z}akowicz et
al. 2013). Fundamental elements of our method are exposed in Section
3. Its main originality relies also on the simultaneous inversion of
the effective temperature $T_{\rm eff}$, the surface gravity log$g$,
the metallicity [Fe/H] \emph{and} the projected rotational velocity
$v{\rm sin}i$ directly and only from a specific spectral band
extracted from the full range covered by Espadons and Narval.

As compared to alternative methods such as $\chi^{2}$ fitting to a
library of (synthetic) spectra, as done by Munari et al. (2005) for
the analysis of the RAVE survey, for instance, the main advantages of
PCA are in the reduction of dimensionality -- a critical issue when
dealing with high-resolution spectra also covering a very large
bandwith -- which allows a very fast processing of the data, and in
the ``denoising'' of the original data (see e.g., Bailer-Jones et
al. 1998 or Paletou 2012 in another context though). It also differs
from another projection method such as Matisse which uses specific
projection vectors attached, say, to each stellar parameter to be
inverted (Recio-Blanco et al. 2006). In that frame, these vectors are
derived after assuming that they are linear combinations of every
``individual'' belonging to a learning database of synthetic spectra.

In this study we restrict ourselves, on purpose, to a spectral domain
ranging from 500 to 540 nm, that is around the Mg\,{\sc i} b-triplet
lines. The main argument in favour of this spectral domain lies in the
fact that spectral lines of this triplet are excellent surface gravity
indicators (e.g., Cayrel \& Cayrel 1963, Cayrel de Strobel 1969),
log$g$ being at the same time the most difficult parameter to retrieve
from spectral data. It is also a spectral domain devoid of telluric
lines. More recently, Gazzano et al. (2010) performed a convincing
spectral analysis of Flames/Giraffe spectra working with a similar
spectral domain. However, in our study we shall use {\it observed}
spectra as training database for our PCA-based method of inversion.

Preliminary tests, done with the inversion of spectra taken from the
so-called S$^4$N survey (Allende Prieto et al. 2004), are discussed in
Section 4. Finally, we proceed with the inversion of 140 spectra of
FGK stars form {\sc PolarBase} for which fundamental parameters are already
available from the so-called {\sc Spocs} catalogue of Valenti \& Fisher
(2005).

%__________________________________________________________________

\section{Sources of data}

Our reference spectra are taken from the Elodie stellar library
(Prugniel et al. 2007, Prugniel \& Soubiran 2001). They are publicly
available\footnote{see: {\tt http://atlas.obs-hp.fr/elodie/}} and
fully documented. An important information concerns the wavelength
coverage of Elodie spectra, which is about 390--680 nm. This,
unfortunately, prevents us from doing tests in the spectral domain at
the vicinity of the infrared triplet of Ca\,{\sc ii} for instance (see
e.g., Munari 1999 for a detailed case concerning this spectral
range). We used also the high-resolution spectra at ${{\cal R} \sim
  42\,000}$.

First tests of our method were done with stellar spectra coming from
the {\it Spectroscopic Survey of Stars in the Solar Neighbourhood},
aka. S$^4$N survey (Allende Prieto et al. 2004). They are also
publicly\footnote{ {\tt hebe.as.utexas.edu/s4n/} } available and fully
documented. The wavelength coverage is 362--1044 nm (McDonald
Observatory, 2.7-m telescope) or 362--961 nm (La Silla, 1.52-m
telescope) and spectra have resolution ${{\cal R} \sim 50\,000}$.

However, our main purpose is the inversion of stellar parameters from
high-resolution spectra coming from Narval and Espadons
spectropolarimeters. Such data is now available from the \emph{public}
database {\sc PolarBase} (Petit et al. 2014). Narval is a state-of-the-art
spectropolarimeter operating in the 380--1000 nm spectral domain,
with a spectral resolution of 65\,000 in its polarimetric mode. It is
an improved copy, adapted to the 2-m TBL telescope, of the Espadons
spectropolarimeter, in operations since 2004 at the 3.6-m aperture
CFHT telescope.

\begin{figure}
%  \sidecaption
  \includegraphics[width=9. cm,angle=0]{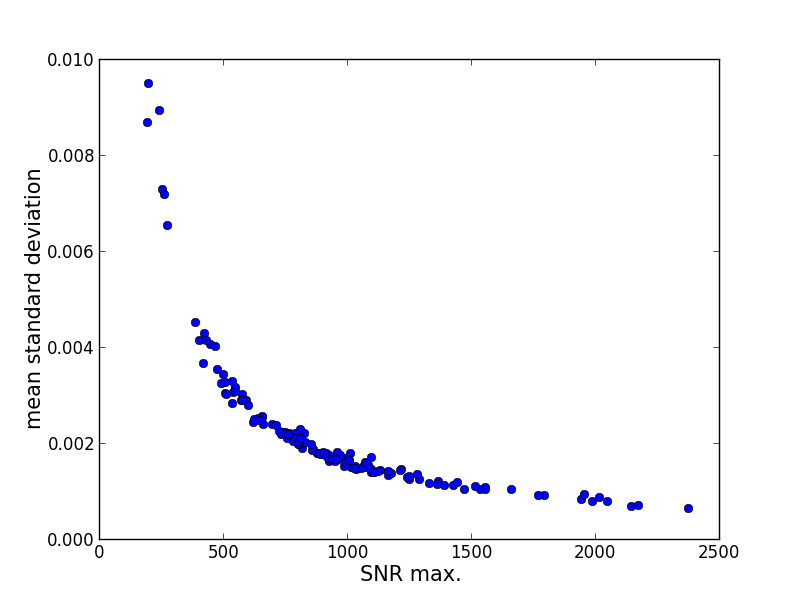}
  \caption{Typical domain of variation of the noise level associated
    with the Espadons-Narval spectra we shall process.  The mean
    standard deviation of noise per pixel, for the wavelength range
    around the b-triplet of Mg\,{\sc i} is displayed here vs. the
    maximum signal to noise ratio of the full spectra.}
  \label{Fig1}
\end{figure}

{\sc PolarBase}  is operational since 2013. It is at the present
time the largest on-line archive of high-resolution polarization
spectra. It hosts data that were taken at the 3.6-m CFHT telescope
since 2005 and with the 2-m TBL telescope since 2006. So far, more than
180\,000 independent spectra are available, for more than 2\,000 distinct
targets all over the Hertzsprung-Russell diagram. More than 30\,000
\emph{polarized} spectra are also available, mostly for
\emph{circular} polarization.  Linear polarization data are very
seldom still and amount to a about 2\% of the available data.

At the present time, the PolarBase database provides no more than
Stokes $I$ or $V /I_c$ spectra calibrated in wavelength. Stokes $I$
data are either normalized to the local continuum or not. We have
however plans to propose higher-level data, such as pseudo-profiles
resulting from line addition and/or least-squares deconvolution (see
e.g., Paletou 2012), activity indexes as well as stellar fundamentals
parameters. The latter's knowledge, besides being obviously
interesting by itself, is also indispensable to any accurate further
post-processing of these high-resolution spectra. These spectra are
also generally bearing high signal-to-noise ratios, as can be seen in
Fig.\,(\ref{Fig1}). Indeed Stokes $I$ spectra we have been using
result from the combination of 4 successive exposures, each of them
carrying 2 spectra of orthogonal polarities generated by a \emph{Savart}
plate-type analyser. This procedure of double so-called
``beam-exchange'' measurement is indeed meant for the purpose of
extracting (very) weak polarization signals (see e.g., Semel et
al. 1993).

%__________________________________________________________________

\section{PCA-based inversion}

Our PCA-based inversion tool is strongly inspired by magnetic and
velocity field inversion tools which have been developed during the
last decade to diagnose solar spectropolarimetric data (see e.g., Rees
et al. 2000). Improvements of this method have been recently exposed
by Casini et al. (2013) for instance. Hereafter, we describe its main
characteristics, in the particular context of our study.

\begin{figure*}
%  \sidecaption
  \includegraphics[width=19cm,angle=0]{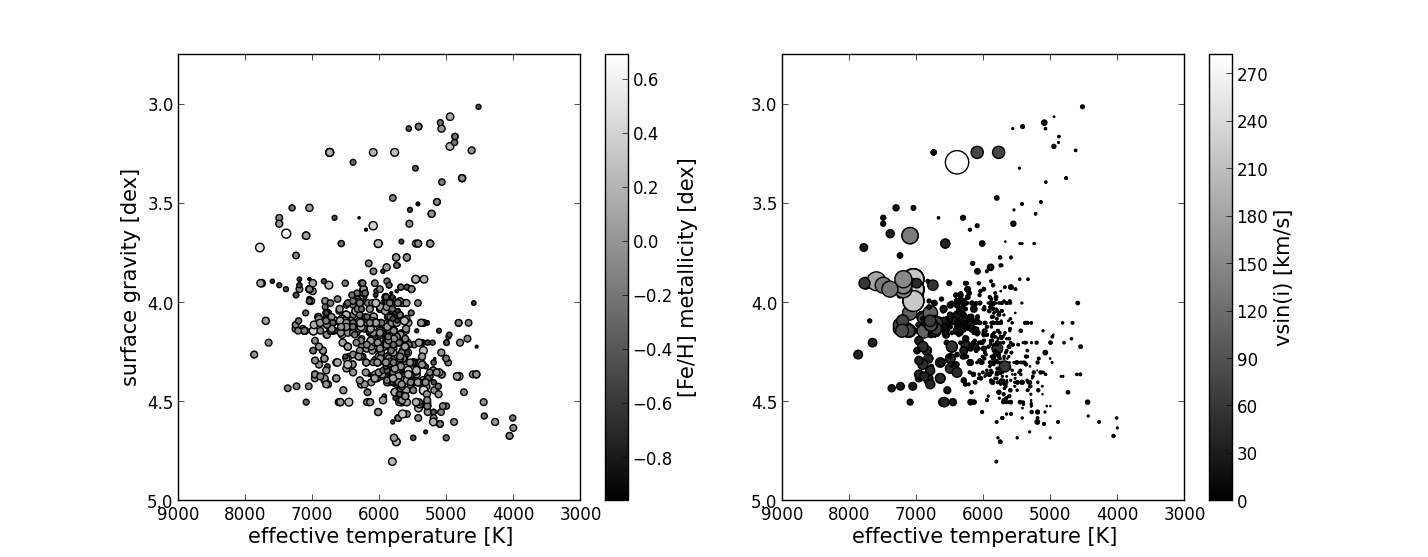}
  \caption{Graphical summary of the coverage in stellar parameters
    corresponding to the content of our Elodie spectra training database
    (for FGK stars). Note that we adopted values of $v{\rm sin}i$ from Vizier@CDS
    catalogue III/244 (G{\l}ebocki \& Gnaci{\'n}ski 2005). Size and color of
    each dot are proportional to either [Fe/H] (left) or $v{\rm sin}i$ (right).}
  \label{Fig2}
\end{figure*}

\subsection{The training database}

Our training database was created after the Elodie stellar spectral
library (Prugniel et al. 2007, Prugniel \& Soubiran 2001; see also
Vizier@CDS catalogue III/251).  Stellar parameters associated with
each spectra have been extracted by us from the CDS, using ressources
from the Python package {\tt astroquery}\footnote{ {\tt
    astroquery.readthedocs.org} }, in particular its components
allowing to query Vizier catalogues.

We had, however, to complement these informations with a value of the
projected rotational velocity for each object/spectra of our
database. To do so, we have been querying the catalogue of stellar
rotational velocities of G{\l}ebocki \& Gnaci{\'n}ski (2005, also Vizier@CDS
catalogue III/244\footnote{We had to correct the value of 25 km\,${\rm
    s}^{-1}$ given for HD\,16232, significantly underestimated
  vs. alternative determinations (with a median of 40 km\,${\rm
    s}^{-1}$)}). Note also that we removed all objects for which we
could not find a value of $v{\rm sin}i$ in this catalogue.

Since our preliminary tests concerns the inversion of the S$^4$N
survey as well as objects common between the content of {\sc
  PolarBase} and the one of the so-called {\sc Spocs} catalogue
(Valenti \& Fisher 2005), we limited ourselves to such spectra for
which T$_{\rm eff}$ lies between 4\,000 and 8\,000 K, log$g$ is
greater than 3.0 dex and [Fe/H] is greater than -1.0 dex. We also had
to reject a few Elodie spectra which we found unproperly corrected for
radial velocity and therefore misaligned in wavelength with respect to
the other spectra. The various coverage in stellar parameters coming
along the finally 905 selected spectra are summarized in
Figs.\,(\ref{Fig2}).

Finally, following Mu\~noz Bermejo et al. (2013; see their \S2.1) we
adopted the same renormalization procedure, homogeneously, for the
whole set of Elodie spectra making our training database. It is an
iterative method consisting in two main steps. The first stage
consists in fitting the normalized flux in the spectral bandwith of
interest by a high-order (8-th) polynomial. Then we compute
$D(\lambda)$ i.e, the difference between the initial spectra and the
polynomial fit, as well as its mean $\bar{D}$ and standard deviation
$\sigma_D$. We reject these wavelengths such that $(D-\bar{D})$ are
either beyond -0.5$\sigma_D$ or above 3$\sigma_D$. This scheme is
iterated 10 times, which guarantees that we properly extract the
continuum envelope of the initial spectra.  Finally, we use the
remaining flux values to renormalize the initial spectra. As
  shows Fig.\,(3), this concerns relatively small corrections of the
continuum level, never exceeding a few percents. As mentioned earlier
by Mu\~noz Bermejo et al. (2013), this procedure may be arguable, but
we found it satisfactory and it was consistently applied to all the
spectra we used.

\begin{figure}
%  \sidecaption
  \includegraphics[width=9cm,angle=0]{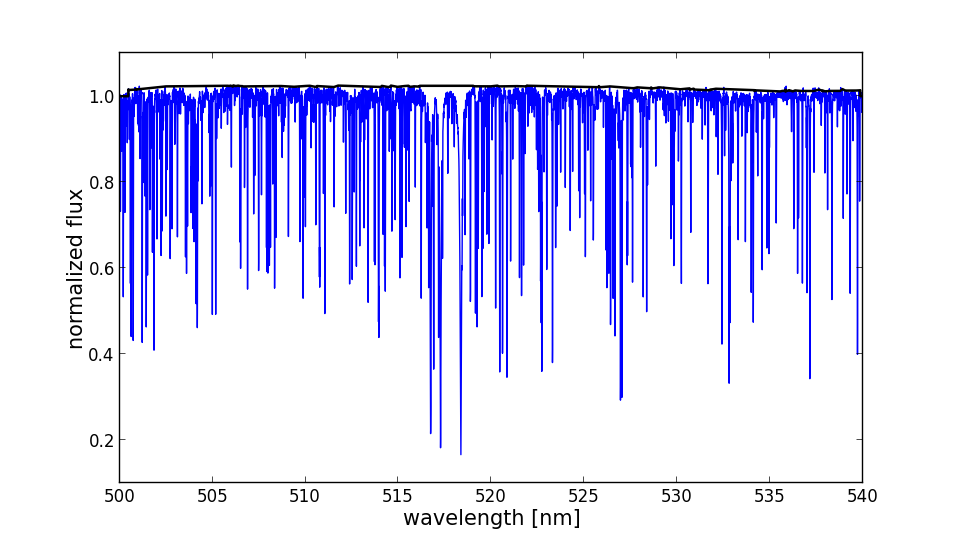}
  \caption{A typical example of the continuum level fit (strong dark
    line) on top of the original spectrum still bearing small
    normalization errors to be corrected for.}
  \label{Fig3}
\end{figure}

\subsection{Reduction of dimensionality}

Espadons and Narval spectropolarimeters typically provide of the order
of 250\,000 flux measurements vs. wavelength across a spectral range
spanning from about 380 to 1000 nm, for each spectra. Hereafter we
shall only consider spectra obtained in the polarimetric mode at a
resolvance of ${{\cal R} \sim 65\,000}$.  Indeed, one of our main
objective is that stellar parameters derived from Stokes $I$ spectra
can be directly used for the further post-processing of the multi-line
\emph{polarized} spectra which is obtained simultaneously (see e.g.,
Paletou 2012 and references therein).

In this study, we adopt a first reduction of dimensionality by
restricting the spectral domain from which we shall invert stellar
parameters to the vicinity of the Mg\,{\sc i} b-triplet that is, for
wavelengths ranging from 500 to 540 nm. The presence of good
surface gravity indicators like the strong lines of the b-triplet of
Mg\,{\sc i} ($\lambda \lambda$ 516.75, 517.25 and 518.36 nm) as well
as several other metallic lines in their neighbourood, and the absence
of telluric lines in this spectral domain are the main arguments we
used. Considering this, the matrix $\vec{S}$ representing our training
database sizes like $N_{\rm spectra}=905$ by $N_{\lambda}=8\,000$.

Next, we compute the eigenvectors {$\vec{e}_k (\lambda)$} of the
variance-covariance matrix defined as

\begin{equation}
\vec{C} = {(\vec{S}- \bar{S})^{T}} \cdot {(\vec{S}- \bar{S})} \, ,
\end{equation} 
where $\bar{S}$ is the mean of $\vec{S}$ along the $N_{\rm
  spectra}$-axis.  Therefore $\vec{C}$ is a $N_{\lambda} \times
N_{\lambda}$ matrix.  In the framework of principal component
analysis, reduction of dimensionality is achieved by representing the
original data by a limited set of projection coefficients

\begin{equation}
{p}_{jk}= ({S}_j - \bar{S}) \cdot \vec{e}_k  \, ,
\end{equation} 
with $k_{\rm max} \ll N_{\lambda}$. In what follows for the
processing of all observed spectra, we shall adopt $k_{\rm max}=12$.

\begin{figure}
%  \sidecaption
  \includegraphics[width=9. cm,angle=0]{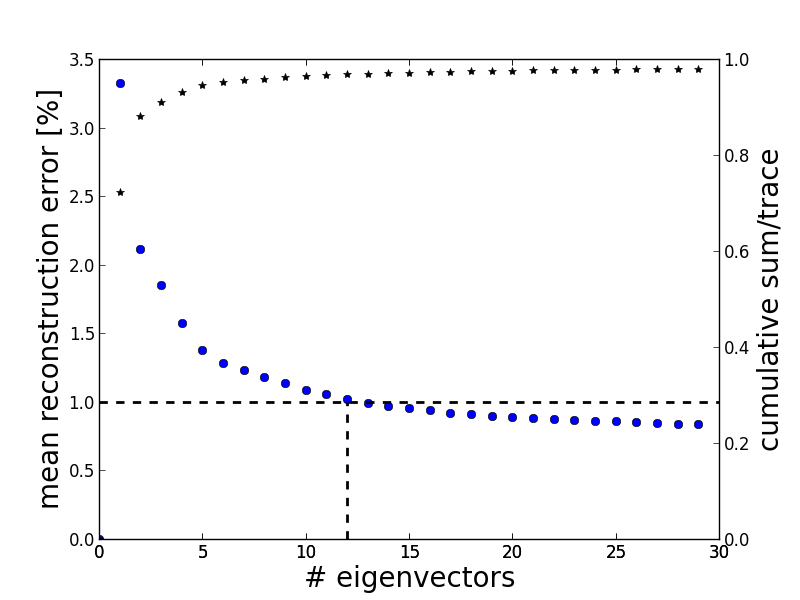}
  \caption{Reconstruction error as a function of the number of
    eigenvectors used for the computation of the projection
    coefficients $p$. The mean reconstruction error gets below 1\%
    (with a standard deviation of 0.4\%) after rank 12. The rightside
    scale is for the cumulative sum of the ordered eigenvalues
    normalized to their sum ($\star$).}
  \label{Fig4}
\end{figure}

The most frequent argument supporting the choice of $k_{\rm max}$
relies on the accuracy achieved for the reconstruction of the original
set of $S_i$'s from a limited set of eigenvectors (see e.g., Rees et
al. 2000 or Mu\~noz Bermejo et al. 2013).  In the present case, we
display in Fig.\,(\ref{Fig4}) the mean reconstruction error

\begin{equation}
  E(k_{\rm max}) = \left \langle \left( {  { \mid\bar{S} + \sum_{k=1}^{k _{\rm max}} { {p}_{jk} \vec{e}_k } -
        {S}_j \mid } \over { S_j }  }  \right) \right \rangle \, ,
\end{equation} 
as a function of the maximum number of eigenvectors considered for the
computation of the projection coefficients $p$. It is noticeable
in Fig. (4) that this reconstruction error is better than 1\% for
$k_{\rm max} \ge 12$. Also, at the same rank, an alternative
  estimator like the cumulative sum of the ordered eigenvalues
  normalized to their sum becomes larger than 0.95.

To conclude this part, we display in Fig.\,(5) the 12
  eigenvectors that we shall use further.

\begin{figure*}
%  \sidecaption
  \includegraphics[width=19. cm,angle=0]{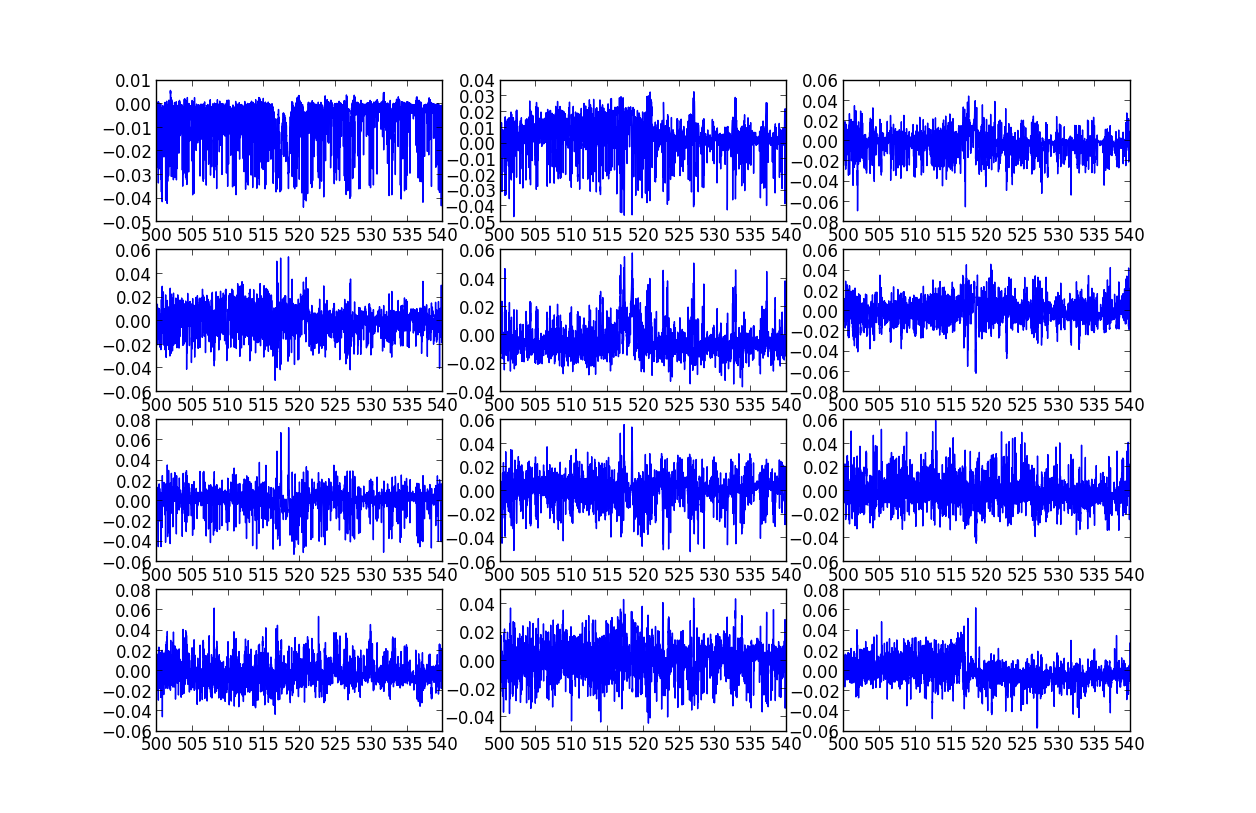}
  \caption{From top left to bottom right, the 12 eigenvectors used in
    our inversion method. Each of them is displayed vs. the same
    wavelength scale in nm.}
  \label{Fig5}
\end{figure*}

\subsection{Nearest neighbour(s) search}

The above described reduction of dimensionality allows one to perform
a fast and reliable inversion of observed spectra, once the latter
have been: \emph{(i)} corrected for the wavelength shift vs. the
spectra in our database, because of the radial velocity of the target,
\emph{(ii)} continuum-renormalized as accurately as possible,
\emph{(iii)} degraded in spectral resolution to be comparable to the
${{\cal R} \sim 42\,000}$ resolvance of the Elodie spectra we use and,
finally \emph{(iv)} resampled in wavelength as the collection of
Elodie spectra. We shall come back to these various stages in the next
section. However once these tasks have been achieved, the inversion
process is the following.

Let $O(\lambda)$ the observed spectrum made comparable to Elodie
ones. We now compute the set of projection coefficients

\begin{equation} {\varrho}_{k}= { { ({O} - \bar{S}) \cdot \vec{e}_{k}
    } }\, .
\end{equation} 
The nearest neighbour search is therefore done
by seeking the minimum of the squared \emph{Euclidian} distance

\begin{equation}
d^{(O)}_{j} = \sum_{k=1}^{k_{\rm max}}  \left( {\varrho}_{k} - {p}_{jk} \right)^{2}\, ,
\end{equation}
where $j$ spans the number, or a limited number if any \emph{a priori}
is known about the target, of distinct reference spectra in the
training database.  In practice, we do not limit ourselves to the
nearest neighbour search, although it already provides a relevant set
of stellar parameters. Because PCA-distances between several
neighbours may be of the same order, we adopted a simple procedure
which consists in considering \emph{all} neighbours in a domain

\begin{equation}
{\rm min} \left ({d}^{(O)}_{j} \right) \leq  {d}^{(O)}_{j} \leq
1.2\times {\rm min} \left( {d}^{(O)}_{j} \right)
\, ,
\label{eq:NNB}
\end{equation} 
and derive stellar parameters as the (simple) mean of each set of
parameters \{$T_{\rm eff}$; log$g$; [Fe/H]; $v{\rm sin}i$\} characterising
this set of nearest neighbours (A. L\'opez Ariste, private
communication). We did not notice significant changes in the results
either for a smaller range of PCA-distances or when adopting e.g.,
distance-weighted mean parameters. This point will be discussed again
in the forthcoming sections.

\subsection{Internal error}

In order to characterize our inversion method, we have been inverting
stellar parameters $T_{\rm eff}$, log$g$, [Fe/H] and $v{\rm sin}i$ for
every spectra (905) constituting the training database. However, at
each step we removed the spectra being processed from the database
(and then recomputed the eigenvalues and eigenvectors of the new
variance-covariance matrix).

\begin{table}
  \caption{Bias and standard deviation of the differences
    between \emph{inverted} Elodie spectra and its initial stellar
    parameters of reference $T_{\rm eff}$ (K), log$g$ and [Fe/H]
    (dex) and $v{\rm sin}i$ (km\,${\rm s}^{-1}$). At
    each step, the Elodie spectra being processed was removed from the
    training database.}
\label{table:1}
\centering
\begin{tabular}{ccccc}
  \hline\hline
  Parameter & $T_{\rm eff}$ & log$g$ & [Fe/H]  & $v{\rm sin}i$ 
  \\
  \hline
  bias & 0.78 & 0.02 & 0.005 & 0.07 \\
  $\sigma$ & 170 & 0.16 & 0.11 & 6.84 \\
  \hline
\end{tabular}
\end{table}

To summarizes this analysis, we give in Table 1 internal errors,
$\sigma$, measured for each inverted stellar parameter. The
disappointing result on $v{\rm sin}i$ mainly comes from a suspicious
scatter of results for these objects having large $v{\rm sin}i$,
typically beyond 100 km\,${\rm s}^{-1}$. However, for our tests with
S$^4$N and {\sc PolarBase} data, we did not have to deal with objects
having $v{\rm sin}i$ beyond 80 km\,${\rm s}^{-1}$ (see next sections).

\section{Inversion of S$^4$N spectra}

A convincing test of our method would be to invert high-resolution
spectra which were gathered in the frame of the so-called {\it
  Spectroscopic Survey of Stars in the Solar Neighbourhood}
(aka. S$^4$N, Allende Prieto et al. 2004)

We selected only these S$^4$N spectra for which \emph{all} three parameters
\{$T_{\rm eff}$; log$g$; [Fe/H]\} have been determined by Allende
Prieto et al. (2004). We used the same reference catalogue as the one
used for the Elodie spectra to add a $v{\rm sin}i$ value to each spectra
(see e.g., Paletou \& Zolotukhin 2014). Since they were acquired at a
higher spectral resolution than Elodie ones, we adapted each spectra
to the latter resolvance ${{\cal R} \sim 42\,000}$ using an
appropriate Gaussian filter. Finally we applied to all spectra the
same renormalization procedure as the one also applied to our Elodie
spectra database.

We identified 49 objects in common between S$^4$N and our sample of
905 objects taken from the Elodie Stellar Library. Using respective
catalogue values for effective temperature $T_{\rm eff}$, surface
gravity log$g$ and metallicity [Fe/H] we could easily estimate bias
and standard deviations between the two distinct estimate, for each
stellar parameter. Results are summarized in Table 2.

\begin{table}
  \caption{Bias and standard deviation of the absolute differences
    between \emph{reference} stellar parameters  $T_{\rm eff}$ (K),
    log$g$ (dex) 
    and [Fe/H]  (dex)  retrieved respectively from the S$^4$N and the Elodie
    catalogues, for the 49 objects in common between our data samples.}
\label{table:2}
\centering
\begin{tabular}{cccc}
  \hline\hline
  Parameter  & $T_{\rm eff}$ & log$g$  & [Fe/H]  \\
  \hline
  bias & -66 & 0.16 & 0.006  \\
  $\sigma$ & 84 & 0.14 & 0.08  \\
  \hline
\end{tabular}
\end{table}

\begin{figure*}
%  \sidecaption
  \includegraphics[width=19cm,angle=0]{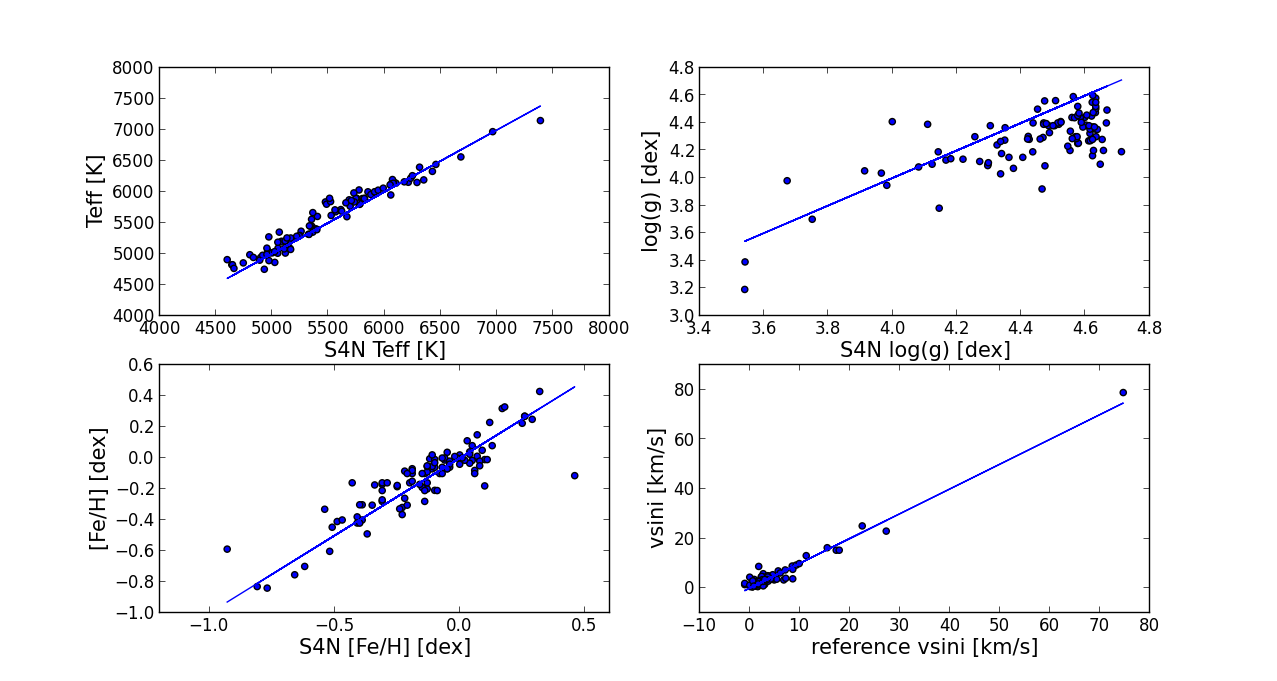}
  \caption{Stellar fundamental parameters inverted from the analysis
    of S$^4$N spectra, using our PCA-Elodie method vs. their S$^4$N
    catalogue values of reference (except for $v{\rm sin}i$ -- see
    text). Respective bias and standard deviation deduced from the
    analysis of the difference between inverted and
    catalogue/reference parameters are: (-74 ; 115) for $T_{\rm eff}$,
    (0.16 ; 0.16) for log$g$ and (-0.01 ; 0.11) for [Fe/H]. For $v{\rm
      sin}i$ values, for which the source of data is not the S$^4$N
    catalogue however, we find (-0.36 ; 1.61). }
  \label{Fig6}
\end{figure*}

This should be now compared to the results of our \emph{inversion} of
104 S$^4$N spectra using our Elodie training database for PCA. They
are summarized, for each stellar parameters including $v{\rm sin}i$
though, both in Fig.\,(\ref{Fig6}) and in Table 3 (where we also
  detail specific values for, respectively, the inversion of the
  spectra of the 49 objects in common between S$^4$N and our Elodie
  sample, and the 55 remaining objects). Bias and dispersions,
especially conspicuous for log$g$, measured after our inverted
parameters appear to be quite direct imprints of the discrepancies
already put in evidence, by the direct comparison between respective
reference values for objects in common between the two samples.
However, taking into account that our inversion relies \emph{only} on
spectroscopic information, moreover on a limited (but relevant)
spectral band, we find our approach satisfactory for our purpose.

\begin{table}
  \caption{Bias and standard deviation of the absolute differences
    between \emph{inverted} and reference stellar parameters $T_{\rm eff}$ (K), log$g$ and [Fe/H]
    (dex) and $v{\rm sin}i$ (km\,${\rm s}^{-1}$), for our sample
    of 104 S$^4$N spectra -- see also Fig.\,(\ref{Fig6}). }
\label{table:3}
\centering
\begin{tabular}{ccccc}
  \hline\hline
  Parameter & $T_{\rm eff}$ & log$g$ & [Fe/H]  & $v{\rm sin}i$ \\
  \hline
  bias & -74 & 0.16 & -0.01 & -0.36 \\
  $\sigma$ & 115 & 0.16 & 0.11 & 1.61 \\
  \hline
  49 objects in common & & & & \\
\hline
  bias & -62 & 0.16 & 0.001 & -0.16 \\
  $\sigma$ & 82 & 0.16 & 0.08 & 0.67 \\
  \hline
  \hline
  55 objects not in common & & & & \\
\hline
  bias & -85 & 0.15 & -0.02 & -0.55 \\
  $\sigma$ & 138 & 0.17 & 0.13 & 2.10 \\
  \hline
\end{tabular}
\end{table}

Our method allows also a reliable and \emph{direct} inversion of the
projected rotational velocity $v{\rm sin}i$ of stars, without any
other limitation than the one coming from the limit in $v{\rm sin}i$
attached to the identified values for spectra present in our Elodie
training database -- see e.g., Figs.\,(2). Using synthetic spectra
(computed for hypothetical non-rotating stars) Gazzano et al. (2010)
study was, for instance, limited to stars having a $v{\rm sin}i$ lower
than about 11 km\,${\rm s}^{-1}$. In our case, we retrieve accurately
$v{\rm sin}i$ values up the most extreme case of $\sim$ 80 ${\rm
  km}\,{\rm s}^{-1}$ in our sample i.e., the high-proper motion F1V
star HIP\,77952\footnote{Note that for this object, one can identify 5
  estimates of $v{\rm sin}i$ from Vizier@CDS ranging from 70 to 90
  ${\rm km}\,{\rm s}^{-1}$, but whose median value of 75 ${\rm
    km}\,{\rm s}^{-1}$ agrees very well with our own estimate.}  i.e., fast
rotators for which usual synthetic models are unsatisfactory.

Another advantage of our method is that nearest neighbour(s) are also
identifiable as \emph{objects} i.e., other stars. In that sense, our
method can also be seen as relevant from \emph{classification}. For
instance, considering the only objects in common between our S$^4$N
and our Elodie spectra/objects sample, in 75\% of the cases the very
nearest neighbour is another spectra of the \emph{same} object and,
for the remainder, conditions expressed in Eq.\,(6) guarantee that the
same object spectra belongs to the set of nearest neighbours.

\section{Inversion of PolarBase spectra}

Hereafter we shall first discuss details about which kind of
conditioning have to be applied to {\sc PolarBase} spectra in order to
be made comparable to our Elodie-based training database. Then we
shall discuss the inversion of spectra of the Sun taken on reflection
over the Moon surface at the TBL telescope, and finally inversions of
140 spectra from FGK objects in common with the sample studied by
Valenti \& Fischer (2005).

\subsection{Conditioning of PolarBase spectra}

The first and obvious task to perform on \emph{observed} spectra is to
correct for their wavelength shift vs. Elodie spectra which are found
already corrected for radial velocity ($v_{\rm rad}$). The radial
velocity of the target at the time of the observation is deduced from
the centroid, in a velocity space, of the pseudo-profile resulting
from the ``addition'' (see e.g., Paletou 2012) of the three spectral
lines of the Ca\,{\sc ii} infrared triplet whose rest wavelengths are,
respectively, 849.802, 854.209 and 866.214 nm. Note that this can be
done with any other set of spectral line supposed to be \emph{a
  priori} present in the spectra we want to process. One of the
advantages of the Ca\,{\sc ii} infrared triplet is its ``persistence''
for spectral types ranging from A to M.  Once $v_{\rm rad}$ is known,
the observed profile is set on a new wavelength grid, at rest.

We could check, using a solar spectra (see also the next section) that
$v_{\rm rad}$ should be known to an accuracy of the order of $\delta
v/4$ with $\delta v=c/{\cal R}$ i.e., about 1.15 ${\rm km}\,{\rm
  s}^{-1}$ with our Espadons-Narval data. Beyond this value, estimates
of $T_{\rm eff}$ and $v{\rm sin}i$ first, start to be significantly
affected by the misalignement of the observed spectral lines with
those of the spectra of the training database. Indeed, the
neighbourhood identified by our PCA-based approach can change quite
dramatically because of such a spectral misalignement.

A second step consists in adapting the resolution of the initial
spectra, about ${{\cal R} \sim 65\,000}$ in the polarimetric mode, to
the one of the training database spectra i.e., ${{\cal R} \sim
  42\,000}$. This is done by convolving the initial observed profile
by a Gaussian profile of adequate width. Then we resample the
wavelength grid down to the one common to all reference spectra, and
we interpolate the original spectra onto the new wavelength
grid. Finally, we apply the same renormalization procedure as already
decribed in \S 3.1, for consistency.

\begin{figure*}
%  \sidecaption
  \includegraphics[width=19cm,angle=0]{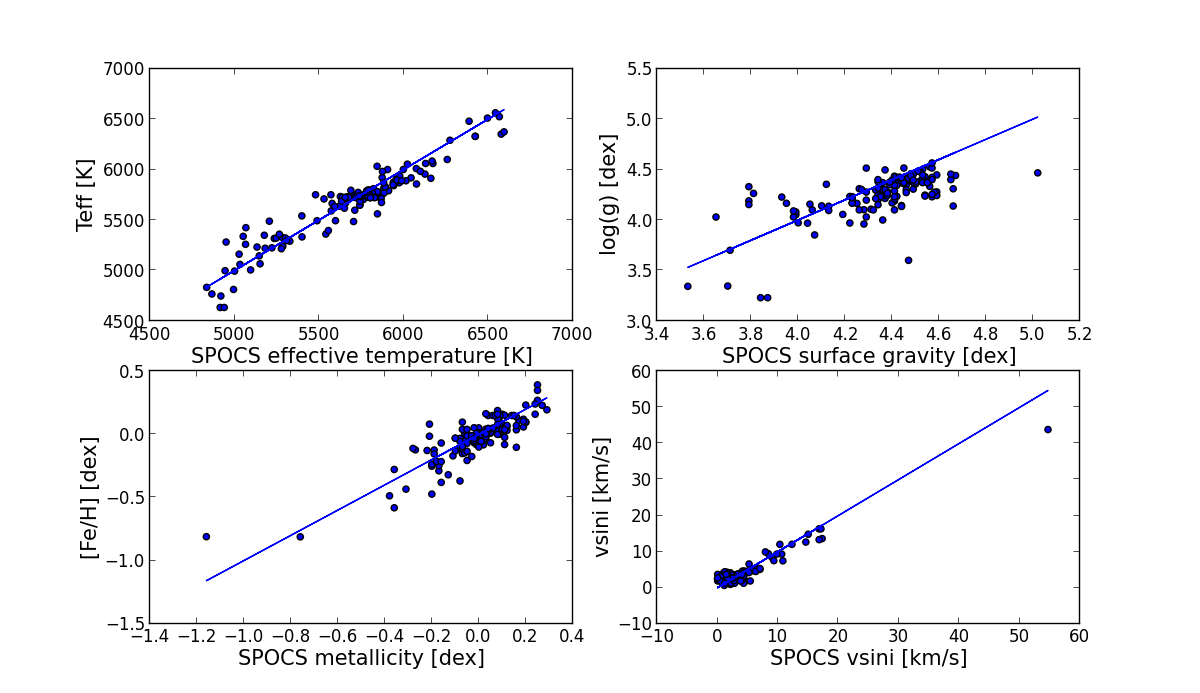}
  \caption{Stellar fundamental parameters inverted from {\sc
      PolarBase} spectra in common with the {\sc Spocs} catalogue,
    using our PCA-Elodie method vs. {\sc Spocs} reference values,
    including $v{\rm sin}i$. Respective bias and standard deviation
    deduced from the analysis of the difference between inverted and
    reference parameters are: (23 ; 115) for $T_{\rm eff}$,
    (0.10 ; 0.19) for log$g$, (0.02 ; 0.10) for [Fe/H], and (-0.04 ;
    1.68) for $v{\rm sin}i$. }
  \label{Fig7}
\end{figure*}

\subsection{Solar spectra observed with Narval}

First tests of our inversion method with {\sc PolarBase} data
were performed using solar spectra observed by the 2-m aperture TBL
telescope by reflection over the surface of the Moon in March and June
2012.

Using the very same training database as the one used for the tests
done with S$^4$N data, we could identify as ``nearest neighbour'' star
to the Sun HD\,186427 (aka. 16 Cyg B). It is indeed a G3V
planet-hosting star often identified as being a solar twin (see e.g.,
Porto de Mello et al. 2014). Its stellar fundamental parameters are:
$T_{\rm eff}$=5757 K, log$g$=4.35 dex, [Fe/H]=0.06 and $v{\rm sin}i
\sim 2.18$ km\,${\rm s}^{-1}$ (see also Tucci Maia et al. 2014, for a
recent determination of these parameters). The other nearest neighbour
we could identify is HD\,29150, a star whose main parameters are:
$T_{\rm eff}$=5733 K, log$g \sim 4.35$ dex (Lee et al. 2011, from
Simbad@CDS query), [Fe/H]=0.0 and $v{\rm sin}i \sim 1.8$ km\,${\rm
  s}^{-1}$.

Taking into consideration this neighbourhood, in the PCA-sense, we thus
derive quite satisfactory estimates for effective temperature $T_{\rm
  eff}$=5745 K, surface gravity log$g$=4.35 dex, a metallicity of
[Fe/H]=0.03 and $v{\rm sin}i \sim 2$ km\,${\rm s}^{-1}$ typical of the
(very) slowly rotating Sun, as observed by reflection over the Moon surface
with the Narval@TBL spectropolarimeter.

This is quite consistent with the test consisting in inverting the Sun
spectra taken from the Elodie stellar library, but \emph{not} a member of our
training database. In such a case, we recover neighbours 16 Cyg B and
HD\,29150 plus the additional HD\,146233 (aka. 18 Sco) and HD\,42807,
a RS\,CVn star of G2V type, also very similar to the Sun indeed.

\begin{table}
  \caption{Bias and standard deviation of the absolute differences
    between stellar parameters  $T_{\rm eff}$ (K), surface
    gravity log$g$ and metallicity [Fe/H] (dex) and projected rotational
    velocity $v{\rm sin}i$ 
    (km\,${\rm s}^{-1}$)
    obtained respectively from
    the {\sc Spocs} catalogue of Valenti \& Fischer (2005, reference values) and our inversion method using Elodie spectra.}
\label{table:4}
\centering
\begin{tabular}{ccccc}
  \hline\hline
  Parameter & $T_{\rm eff}$ & log$g$ & [Fe/H] & $v{\rm sin}i$ \\
  \hline
  bias & 23 & 0.10 & 0.02 & -0.04 \\
  $\sigma$ & 115 & 0.19 & 0.10 & 1.68 \\
  \hline
\end{tabular}
\end{table}

\subsection{Other FGK stars}

We identified in the present content of {\sc PolarBase} 140 targets
which are also identified in the {\sc Spocs} catalogue. For our next
tests of our method, we selected these spectra having the best signal
to noise ratio for every object. Typical values were already given in
Fig.\,(\ref{Fig1}).

Results and characterization of our inversion are given both in
Table\,{\ref{table:4}} and, with more details in
Fig.\,(\ref{Fig7}). The latter figure also gives an idea about the
range of variations of parameters which are expected for the set of
spectra/objects being studied here.  Overall figures are quite similar
to the ones obtained with the S$^4$N although bias values for T$_{\rm
  eff}$ and log$g$ are smaller with {\sc Spocs} data.

\subsubsection{Effective temperature}

Ths standard deviation on the differences between our values and {\sc
  Spocs} reference ones is the same as the one we could evaluate using
S$^4$N spectra (see. Table 3). However the bias value of 23 K is much
smaller for this sample of objects and {\sc PolarBase} spectra. We
note also that the most important dispersion is for the coolest
objects of our sample, for T$_{\rm eff}$ about 5000 K.

A more detailed inspection of outliers in effective temperature,
taking into account alternative and more recent estimates of T$_{\rm
  eff}$ than the one adopted from {\sc Spocs}, reveals that the most
extreme $\Delta$T$_{\rm eff}$ we identified are quite often
overestimated. This is for instance the case for the K1V star LHS\,44
for which a recent determination by Maldonado et al. (2012) is +200 K
from the one of Valenti \& Fischer (2005) and only 140 K from
ours. Another effect may also come from new estimates of parameters
for Elodie objects themselves, an issue we shall discuss hereafter.

Our estimates of effective temperature from spectropolarimetric data
are however already satisfactory for being used in the selection of a
proper ``mask'' (i.e., at least a list of spectral lines) that will
be, in turn, used for the further extraction of polarized signatures.

\subsubsection{Surface gravity}

It is well-known that surface gravity is the most difficult parameter
to get from the analysis of spectroscopic data.  The most conspicuous
outliers show quite easily on the log$g$ subplot in
Fig.\,(\ref{Fig7}). In the ``top-3'' of them showing $\Delta {\rm
  log}g \sim 0.5$ dex or above, we find: EK Dra, LTT\,8785 and
HD\,22918. Concerning EK Dra, our estimate of log$g \sim 3.6$ dex is
way too low as compared to (seldom) values found in the litterature
($\sim 4.5$ dex). This may be compensated by the fact that its
identified nearest neighbour star is the G2IV \emph{subgiant}
HD\,126868 (aka. 105 Vir) for which log$g=3.6$ dex in the Elodie
catalogue, although a value of 3.9 dex is reported elsewhere (see
e.g., the {\sc Pastel} catalogue: Soubiran et al. 2010). For LTT\,8785 and
HD\,22918, examination of alternative estimates for log$g$ (e.g.,
Massarotti et al. 2008; Jones et al. 2011) show that {\sc Spocs}
values may have been slightly overestimated. Besides, the surface
gravity (log$g \sim 3.23$) of the nearest neighbour (and the same
object in both cases) HD\,42983 may have been underestimated. An
inspection of Vizier@CDS for this object indicates 4 different
estimates ranging from 3.23 to 3.6 with a median value of 3.5. Taking
this into account, $\Delta {\rm log}g$ does not exceed 0.2 dex between
our inverted values and reference ones, which is satisfactory.

\subsubsection{Metallicity}

Let us now inspect our results for metallicity. First of all, we
included into our sample LHS\,44 which [Fe/H]=-1.16 according to
Valenti \& Fischer (2005), a value \emph{a priori} excluded from our
working range. But even though, our inversion method points at these
objects in our sample bearing the lowest [Fe/H] values. Then by
decreasing order of $\Delta$[Fe/H] between our inverted values and
reference values, we find: HD\,30508, 40 Eri and LHS\,3976. We found
systematically better agreement between our estimate and statistics on
\emph{all} data available at Vizier, to within 0.05 dex.

\subsubsection{Projected rotational velocity}

Our determinations of $v{\rm sin}i$ are correct and especially
interesting for those objects having a significant projected
rotational velocity, say greater than about 10 km\,${\rm s}^{-1}$. The
major outlier, as seen in Fig\,(\ref{Fig7}), is HR\,1817 for which we
derive a $v{\rm sin}i \sim$ 43 km\,${\rm s}^{-1}$ while {\sc Spocs}
value is about 55 km\,${\rm s}^{-1}$. It is a F8V RS\,CVn star aka. AF
Lep for which another value of 52.6 km\,${\rm s}^{-1}$, still about 10
km\,${\rm s}^{-1}$ greater than our estimate, was more recently
published by Schroeder et al. (2009).

Concerning the determination of $v{\rm sin}i$, it is true that other
methods of evaluation already exist. However, to the best of our
knowledge, they require a template (synthetic) spectrum at $v{\rm
  sin}i\sim 0$ or, at least, a list of spectral lines \emph{a priori}
expected in the spectra, as auxilliary and ``support'' data (see e.g.,
D\'{\i}az et al. 2012 and references therein). Data processing tools
that we shall attach further to {\sc PolarBase} will therefore include
a complementary Fourier analysis module providing an additional $v{\rm
  sin}i$ determination, once stellar fundamental parameters will have
been available from our inversion tool\footnote{The same is true for
  the refinement of radial velocity measurements which can be improved
  by ``line addition'' once identified a proper list of expected
  spectral line wavelengths (at rest).}. Note also that, with our
PCA-based method, we are mostly interested in the ``intermediate''
$v{\rm sin}i$ regime, say between 10 and 100 km\,${\rm
  s}^{-1}$. Indeed, for slower rotators for which rotational
broadening becomes of the order of other sources of broadening (e.g.,
instrumental or turbulent), a more detailed or specific line profile
analysis may be required.

\section{Discussion}

As briefly remarked earlier, the fact that the current implementation
of our method relies on observed data makes it also somewhat relevant
to \emph{classification}. Indeed, we do not just identify ``nearest
spectra'' since nearest neighbour(s) can also be identified as
\emph{objects} i.e., other stars. This important fact is also totally
\emph{independent} from whatever method of determination of stellar
parameters have been used for these objects.

Therefore, unless modifying the sample of spectra/objects constituting
our training database, we do not expect any change in the relation
between inverted spectra and nearest neighbours \emph{as objects},
even though evaluations of their various stellar parameters still may
change in time. Another interesting point is that, this is also true
for \emph{any other stellar parameter} -- especially those
contributing to the spectral signature of a star, beyond the limited
set of fundamental parameters we considered in this study.

%
%--- warning sizes critical for positioning of last longtable!
%
\begin{figure}[]
%  \sidecaption
  \includegraphics[width=7.25 cm,angle=0]{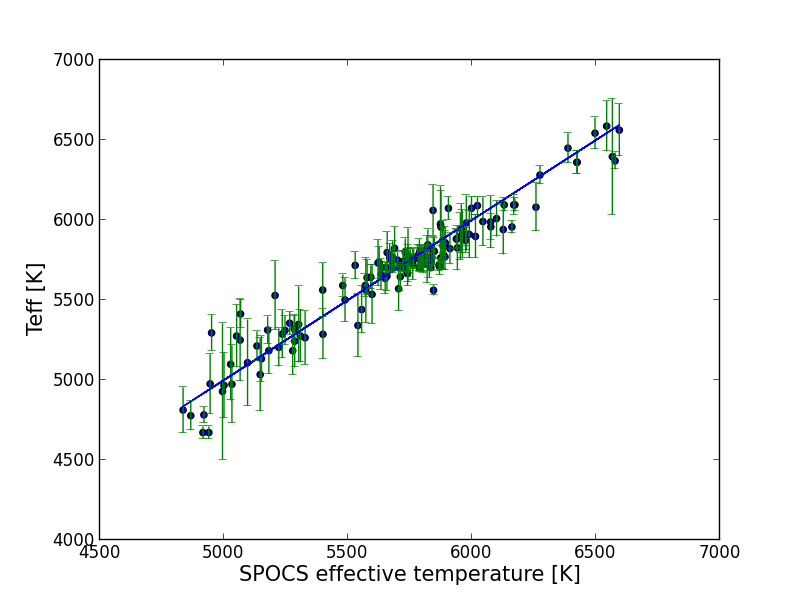}
  \includegraphics[width=7.25 cm,angle=0]{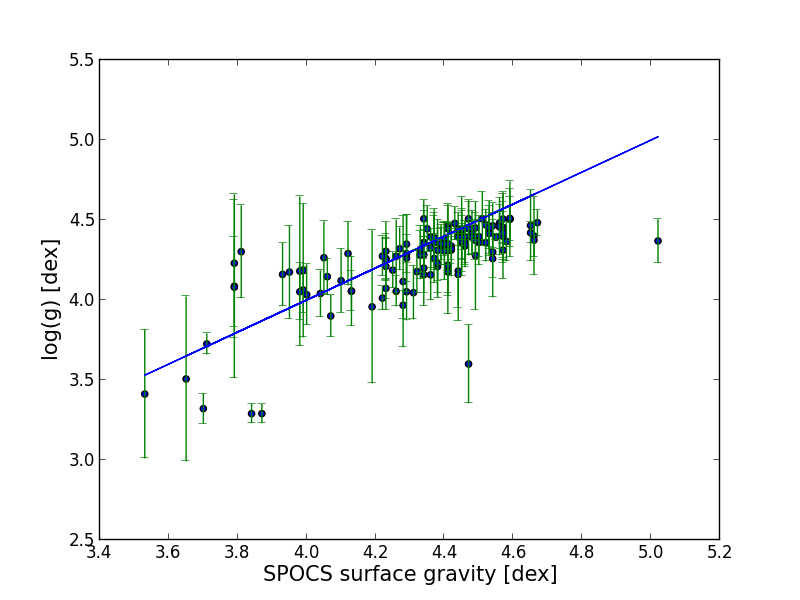}
  \includegraphics[width=7.25 cm,angle=0]{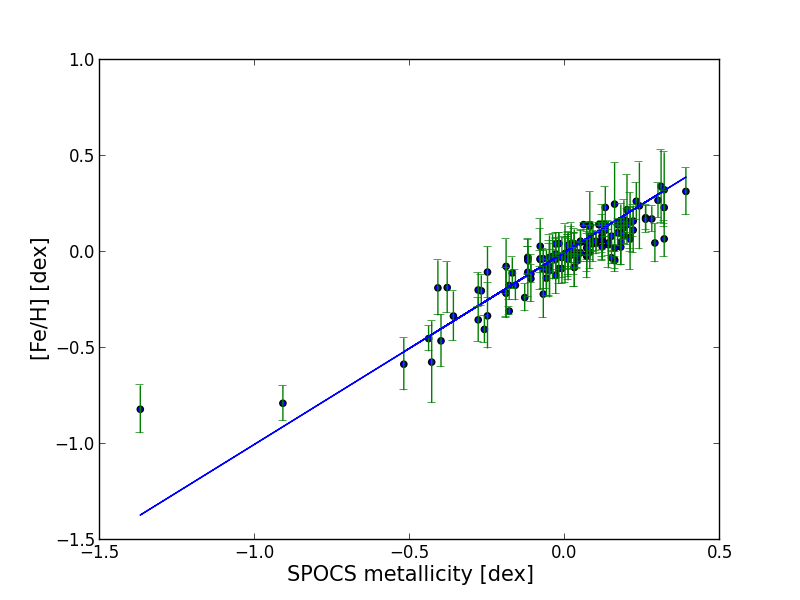}
  \caption{Stellar fundamental parameters, and their respective
    uncertainties, inverted from {\sc PolarBase} spectra in common
    with the {\sc Spocs} catalogue, using our PCA-Elodie method
    vs. {\sc Spocs} reference values. These values were deduced from
    all determinations for each nearest neighbour compiled in the {\sc
      Pastel} catalogue of Soubiran et al. (2010).}
  \label{Fig8}
\end{figure}

Taking advantage of this, instead of using values given by the only
Elodie catalogue, for each {\sc Spocs} spectra we analysed, we have
gathered for every nearest neighbour \emph{all} the evaluations of
effective temperature, surface gravity and metallicity provided by the
comprehensive {\sc Pastel} catalogue\footnote{For 4 targets for which
  we could not retrieve data from {\sc Pastel}, we used instead its
  TGMET values (Katz et al. 1998) also available from the Elodie@Vizier
  catalogue.}  (Soubiran et al. 2010). We could also evaluate
uncertainties on each fundamental parameters as the standard deviation
of the full set of collected values, except the unavailable projected
rotational velocity though. Results are displayed in
Figs.\,(\ref{Fig8}) and details are given in Table\,5.  Mean errors
given in Table\,5 are, respectively, 110 K, 0.16 dex and 0.09 dex for
$T_{\rm eff}$, log$g$ and [Fe/H] i.e., quite consistent, but slightly
better than standard deviation values already given in Table\,4.

As a final remark, it is also worth mentioning that, although we used
it together with a training database made of observed spectra, our
PCA-based inversion method can be equally implemented using
\emph{synthetic} spectra.

\section{Conclusion}

We have implemented a fast and reliable PCA-based numerical method for
the inversion of stellar fundamental parameters $T_{\rm eff}$, log$g$
and [Fe/H], as well as the projected rotational velocity $v{\rm
  sin}i$, from the analysis of high-resolution Echelle spectra
delivered by Narval and Espadons spectropolarimeters. First tests,
mainly made with FGK-stars spectra, show good agreement between our
inverted stellar parameters and reference values published by Allende
Prieto et al. (2004) and Valenti \& Fischer (2005).  We also believe
that our method will be efficient as well for the analysis of
spectra from cooler M stars, as well as from hotter stars, up to
spectral type A.

We used it, so far, with a spectral band located at the vicinity of the
b-triplet of Mg\,{\sc i} and without any help from additional (e.g.,
photometric) information, which is particularly challenging. However
we can easily, either extend or change the spectral domain of use, or
combine analyses from several distinct spectral domains, in order to
constrain further and refine stellar parameters determination. In that
respect, it is important to realize that PCA allows for a quite
dramatic reduction of dimensionality, of the order of 800 ($\sim
N_{\lambda}/k_{\rm max}$) for the configuration we presented here. This
capability is indeed of great interest for a comprehensive
post-processing of high-resolution spectra covering a very large
bandwidth like the ones from Espadons and Narval spectropolarimeters.

\begin{acknowledgements}
  This research has made use of the VizieR catalogue access tool, CDS,
  Strasbourg, France. The original description of the VizieR service
  was published in A\&AS 143, 23. This research has made use of the
  SIMBAD database, operated at CDS, Strasbourg, France. {\sc
    PolarBase} data were provided by the OV-GSO ({\tt
    ov-gso.irap.omp.eu}) datacenter operated by CNRS/INSU and the
  \emph{Universit\'e Paul Sabatier, Observatoire Midi-Pyr\'en\'ees},
  Toulouse (France).
\end{acknowledgements}

%%  \bibliographystyle{aa}
%%  \bibliography{MyBib.bib}

\onecolumn

%\onltab{
%%\begin{longtab}
%\label{table:5}
%%\centering
\begin{longtable}{ccrcrrr}
  \caption{Values and respective uncertainties of the inverted stellar
    parameters $T_{\rm eff}$, log$g$ and [Fe/H] for {\sc Spocs}
    objects in common with {\sc
      PolarBase} using our Elodie based PCA method.} \\
  \hline\hline Object & $T_{\rm eff}$ [K]& $\sigma_{T_{\rm eff}}$ [K]
  & log$g$ [dex]& $\sigma_{{\rm log}g}$ [dex] & [Fe/H] [dex] &
  $\sigma_{\rm
    [Fe/H]}$ [dex] \\
  \hline
\endfirsthead
\caption{continued.}\\
\hline\hline
  Object & $T_{\rm eff}$ [K]& $\sigma_{T_{\rm eff}}$ [K] & log$g$
  [dex]& $\sigma_{{\rm log}g}$ [dex] & [Fe/H] [dex] & $\sigma_{\rm
    [Fe/H]}$ [dex] \\
\hline
\endhead 
\hline
\endfoot
HR753 &  4777 &   91 & 4.40 & 0.24 & -0.03 &  0.09 \\
LTT11292 &  5562 &  203 & 4.46 & 0.14 & -0.02 &  0.11 \\
*39Tau &  5763 &   75 & 4.36 & 0.12 &  0.05 &  0.06  \\
LTT11169 &  5098 &  226 & 3.51 & 0.52 & -0.10 &  0.13  \\
LTT11282 &  5990 &  153 & 4.21 & 0.20 &  0.04 &  0.10  \\
*109Psc &  5768 &  100 & 4.29 & 0.20 &  0.17 &  0.10  \\
*107Psc &  5183 &  145 & 4.45 & 0.17 & -0.03 &  0.12  \\
LHS5051a &  5645 &   86 & 4.40 & 0.03 &  0.13 &  0.03  \\
HD12846 &  5734 &  100 & 4.31 & 0.16 & -0.20 &  0.09  \\
LTT10989 &  5702 &   19 & 4.39 & 0.02 &  0.14 &  0.01  \\
*13Tri &  5854 &   77 & 3.90 & 0.13 & -0.24 &  0.07  \\
HD30825 &  5312 &   85 & 3.73 & 0.07 & -0.11 &  0.08  \\
LHS1753 &  5821 &   98 & 4.26 & 0.15 &  0.03 &  0.07  \\
G175-33 &  5724 &  120 & 4.21 & 0.17 &  0.23 &  0.09  \\
HD9472 &  5718 &   39 & 4.48 & 0.08 & -0.00 &  0.05  \\
V*V451And &  5732 &  144 & 4.47 & 0.10 & -0.00 &  0.08  \\
LTT10580 &  5702 &   19 & 4.39 & 0.02 &  0.14 &  0.01  \\
LHS1125 &  4974 &  242 & 4.51 & 0.14 &  0.05 &  0.12  \\
HD5065 &  5945 &  154 & 4.05 & 0.22 & -0.10 &  0.14  \\
*etaCas &  5825 &  140 & 4.18 & 0.23 & -0.33 &  0.17  \\
LTT11104 &  6369 &   53 & 3.97 & 0.09 &  0.08 &  0.06  \\
V*V987Cas &  5264 &  168 & 4.46 & 0.14 &  0.00 &  0.12  \\
G245-27 &  5649 &   91 & 4.42 & 0.14 &  0.04 &  0.09  \\
HD73344 &  6095 &   41 & 4.18 & 0.13 &  0.11 &  0.04  \\
*55Cnc &  5287 &  151 & 4.43 & 0.09 &  0.34 &  0.19  \\
LTT12401 &  5897 &  134 & 4.30 & 0.16 & -0.00 &  0.09  \\
HR3499 &  6080 &  150 & 4.18 & 0.27 &  0.04 &  0.12  \\
*24LMi &  5768 &   83 & 4.26 & 0.13 &  0.05 &  0.05  \\
LHS2216 &  5749 &   50 & 4.35 & 0.18 &  0.16 &  0.06  \\
HR4486 &  5840 &  154 & 4.36 & 0.17 & -0.21 &  0.13  \\
*36UMa &  5940 &  156 & 4.20 & 0.24 & -0.22 &  0.12  \\
HD98618 &  5773 &   57 & 4.34 & 0.11 &  0.04 &  0.05  \\
V*V377Gem &  5781 &  168 & 4.48 & 0.07 & -0.02 &  0.08  \\
NLTT17627 &  5804 &   86 & 4.34 & 0.12 & -0.01 &  0.09  \\
LTT12204 &  5822 &   81 & 4.34 & 0.13 & -0.02 &  0.07  \\
HD71881 &  5845 &   96 & 4.26 & 0.15 & -0.03 &  0.08  \\
HD138573 &  5739 &   65 & 4.36 & 0.11 &  0.04 &  0.06  \\
HR5659 &  5702 &   19 & 4.39 & 0.02 &  0.14 &  0.01  \\
GJ566A &  5591 &   69 & 4.42 & 0.07 & -0.03 &  0.04  \\
HD129814 &  5808 &   88 & 4.31 & 0.13 & -0.01 &  0.08  \\
*tauBoo &  6449 &   93 & 4.27 & 0.23 &  0.26 &  0.10  \\
*sigBoo &  6396 &  362 & 3.96 & 0.48 & -0.57 &  0.21  \\
LHS348 &  5987 &  163 & 4.31 & 0.18 &  0.05 &  0.09  \\
LHS2498 &  5275 &  195 & 4.09 & 0.57 & -0.19 &  0.13  \\
V*EKDra &  5561 &   32 & 3.60 & 0.24 & -0.04 &  0.02  \\
V*HNPeg &  5872 &   78 & 4.51 & 0.12 & -0.09 &  0.08  \\
HR8455 &  5796 &  129 & 4.17 & 0.29 & -0.11 &  0.13  \\
V*V376Peg &  6009 &  112 & 4.23 & 0.21 & -0.04 &  0.11  \\
HD195019 &  5787 &   96 & 4.25 & 0.13 &  0.05 &  0.07  \\
LTT16813 &  5341 &  198 & 4.40 & 0.25 &  0.07 &  0.16  \\
V*V454And &  5710 &   54 & 4.44 & 0.13 &  0.08 &  0.08  \\
HR8832 &  4812 &  142 & 4.46 & 0.18 &  0.10 &  0.15  \\
LHS544 &  5213 &   92 & 4.39 & 0.19 &  0.00 &  0.10  \\
LTT15779 &  5770 &   69 & 4.32 & 0.12 &  0.06 &  0.05  \\
LTT15881 &  5702 &   19 & 4.39 & 0.02 &  0.14 &  0.01  \\
LTT16028 &  5805 &   33 & 4.39 & 0.15 & -0.11 &  0.05  \\
HD173701 &  5286 &  156 & 4.43 & 0.10 &  0.32 &  0.20  \\
HR7294 &  5745 &   70 & 4.36 & 0.10 &  0.04 &  0.06  \\
LTT15729 &  6562 &  164 & 4.04 & 0.14 & -0.04 &  0.16  \\
*16CygB &  5776 &   70 & 4.28 & 0.11 &  0.06 &  0.05 \\
*sigDra &  5310 &   89 & 4.45 & 0.14 & -0.07 &  0.14  \\
LTT149 &  5754 &   62 & 4.33 & 0.16 &  0.16 &  0.09  \\
LHS146 &  5319 &   81 & 4.51 & 0.24 & -0.58 &  0.14  \\
HD12328 &  4782 &   50 & 3.32 & 0.09 &  0.05 &  0.10  \\
HR448 &  5881 &   57 & 4.06 & 0.12 &  0.18 &  0.07  \\
LTT1267 &  5778 &   64 & 4.27 & 0.12 &  0.10 &  0.07  \\
HD3821 &  5738 &   98 & 4.36 & 0.13 & -0.14 &  0.10  \\
G270-82 &  5634 &   98 & 4.40 & 0.15 & -0.21 &  0.13  \\
LTT709 &  6090 &   57 & 4.19 & 0.11 &  0.05 &  0.06  \\
LTT10409 &  5412 &   90 & 4.30 & 0.29 & -0.19 &  0.14  \\
HR222 &  4976 &  186 & 4.51 & 0.17 & -0.20 &  0.08  \\
HD9986 &  5744 &   66 & 4.36 & 0.11 &  0.04 &  0.06  \\
LTT8887 &  5440 &  148 & 4.28 & 0.34 &  0.15 &  0.15  \\
LTT9062 &  5764 &   87 & 4.32 & 0.15 & -0.03 &  0.10  \\
HR8931 &  5884 &   47 & 4.05 & 0.17 & -0.40 &  0.07  \\
G157-8 &  5897 &  147 & 4.29 & 0.24 & -0.14 &  0.12  \\
HR8734 &  5570 &  137 & 4.26 & 0.24 &  0.32 &  0.12  \\
LTT8785 &  4672 &   38 & 3.29 & 0.06 & -0.05 &  0.05  \\
HD208776 &  5981 &  172 & 4.03 & 0.19 & -0.07 &  0.16  \\
HD377 &  5975 &  236 & 4.12 & 0.41 & -0.03 &  0.03  \\
LHS1239 &  5724 &   99 & 4.35 & 0.14 & -0.17 &  0.08  \\
G131-59 &  5799 &   86 & 4.32 & 0.13 & -0.01 &  0.09  \\
*54Psc &  5204 &  119 & 4.36 & 0.19 &  0.02 &  0.12  \\
HD218687 &  6073 &   72 & 4.45 & 0.14 & -0.04 &  0.04  \\
LTT16778 &  5348 &  238 & 4.18 & 0.42 &  0.22 &  0.18  \\
V*MTPeg &  5783 &   88 & 4.36 & 0.14 &  0.02 &  0.07  \\
*51Peg &  5702 &   19 & 4.39 & 0.02 &  0.14 &  0.01  \\
G131-18 &  5355 &   72 & 4.51 & 0.18 & -0.45 &  0.07  \\
V*V439And &  5640 &  110 & 4.37 & 0.12 &  0.03 &  0.09  \\
V*V344And &  5660 &   96 & 4.37 & 0.10 &  0.07 &  0.09  \\
LHS3976 &  5666 &   79 & 4.31 & 0.18 & -0.46 &  0.14  \\
V*AFLep &  6542 &   99 & 4.37 & 0.14 &  0.05 &  0.10  \\
HR2622 &  5957 &   81 & 4.05 & 0.09 &  0.09 &  0.05  \\
HD46375 &  5243 &  164 & 4.42 & 0.13 &  0.24 &  0.23  \\
*alfCMi &  6587 &  157 & 4.06 & 0.14 & -0.01 &  0.15  \\
LTT2093 &  5882 &   60 & 4.07 & 0.13 &  0.17 &  0.07  \\
*40Eri &  5133 &  142 & 4.45 & 0.22 & -0.35 &  0.11  \\
HD30508 &  5528 &  215 & 4.08 & 0.32 &  0.03 &  0.14  \\
HR1232 &  4928 &  426 & 3.41 & 0.40 &  0.11 &  0.12  \\
LTT1723 &  5249 &  258 & 4.23 & 0.40 &  0.13 &  0.18  \\
LHS1577 &  5823 &  134 & 4.22 & 0.19 & -0.79 &  0.09  \\
HD22918 &  4672 &   38 & 3.29 & 0.06 & -0.05 &  0.05  \\
V*kap01Cet &  5745 &  101 & 4.45 & 0.09 &  0.08 &  0.11  \\
*1Ori &  6361 &   73 & 4.05 & 0.18 & -0.08 &  0.10  \\
HR2251 &  5857 &   99 & 4.28 & 0.15 & -0.02 &  0.08  \\
LTT11933 &  5911 &  151 & 4.35 & 0.20 & -0.09 &  0.10  \\
*37Gem &  5770 &   82 & 4.16 & 0.16 & -0.31 &  0.02  \\
V*chi01Ori &  5872 &   78 & 4.51 & 0.12 & -0.09 &  0.08  \\
LTT5873 &  5275 &  164 & 4.46 & 0.14 & -0.01 &  0.11  \\
LTT3686 &  5702 &   19 & 4.39 & 0.02 &  0.14 &  0.01  \\
LHS2465 &  5956 &   40 & 4.01 & 0.07 &  0.10 &  0.03  \\
LTT12723 &  5751 &   67 & 4.38 & 0.10 &  0.04 &  0.05  \\
*17Vir &  6095 &   41 & 4.18 & 0.13 &  0.11 &  0.04  \\
V*LWCom &  5702 &   19 & 4.39 & 0.02 &  0.14 &  0.01  \\
LTT13145 &  5910 &  146 & 4.31 & 0.19 & -0.03 &  0.11  \\
LTT13442 &  6281 &   57 & 4.15 & 0.11 &  0.17 &  0.07  \\
*61UMa &  5500 &  137 & 4.48 & 0.10 & -0.04 &  0.10  \\
LHS44 &  5294 &  112 & 4.47 & 0.22 & -0.82 &  0.12  \\
LTT7713 &  5762 &   61 & 4.32 & 0.07 &  0.06 &  0.05  \\
HD175726 &  6073 &   72 & 4.45 & 0.14 & -0.04 &  0.04  \\
*18Sco &  5752 &   75 & 4.35 & 0.13 &  0.04 &  0.06  \\
V*V2133Oph &  5183 &  150 & 4.44 & 0.18 & -0.02 &  0.12  \\
V*V2292Oph &  5591 &   69 & 4.42 & 0.07 & -0.03 &  0.04  \\
HD169822 &  5643 &   73 & 4.44 & 0.10 & -0.04 &  0.07  \\
HD159909 &  5733 &   73 & 4.35 & 0.13 &  0.08 &  0.08  \\
HR6950 &  5562 &  169 & 4.16 & 0.20 &  0.23 &  0.11  \\
*110Her &  6361 &   73 & 4.05 & 0.18 & -0.08 &  0.10  \\
LTT15317 &  5747 &   77 & 4.40 & 0.13 & -0.09 &  0.10  \\
HD166435 &  6060 &  160 & 4.16 & 0.29 & -0.02 &  0.07  \\
*86Her &  5535 &  184 & 4.12 & 0.20 &  0.27 &  0.09  \\
HR6806 &  4966 &  205 & 4.45 & 0.17 & -0.17 &  0.08  \\
V*delEri &  5108 &  270 & 4.18 & 0.47 &  0.25 &  0.21  \\
V*epsEri &  5034 &  228 & 4.51 & 0.16 & -0.12 &  0.10  \\
LTT1601 &  5716 &   84 & 4.26 & 0.23 & -0.33 &  0.13  \\
LHS1845 &  5748 &   70 & 4.35 & 0.10 &  0.04 &  0.06  \\
V*V401Hya &  5746 &   63 & 4.39 & 0.11 &  0.08 &  0.08  \\
LTT3283 &  5702 &   19 & 4.39 & 0.02 &  0.14 &  0.01  \\
HD145825 &  5742 &   71 & 4.37 & 0.12 &  0.04 &  0.06  \\
HD143006 &  5956 &  227 & 4.30 & 0.18 &  0.16 &  0.08  \\
HR7291 &  6094 &   60 & 4.16 & 0.11 &  0.05 &  0.07  \\
\end{longtable}
%%\end{longtab}
%} %-end of onltab

\end{document}